\documentclass[aps,prl,twocolumn,superscriptaddress,amsmath,amssymb,reprint,longbibliography]{revtex4-2}
\usepackage{xcolor}
\usepackage{amsmath}
\usepackage{mathtools}
\usepackage{amssymb}
\usepackage{amsfonts}
\usepackage{amsthm}
\usepackage{mathrsfs}
\usepackage{graphicx} 
\usepackage[colorlinks=true
,urlcolor=teal
,anchorcolor=blue
,citecolor=teal
,filecolor=blue
,linkcolor=teal
,menucolor=blue
,linktocpage=true
,pdfusetitle
,pdfauthor={Patrick Hayden, Alexander Maloney, Jinzhao Wang, and Yuxiang Yang}
,pdfa=true
]{hyperref}

\newtheorem{rem}{Remark}

\newcommand{\dd}{\mathrm{d}}
\newcommand{\ket}[1]{| #1 \rangle}

\newcommand{\bra}[1]{\langle #1 |}

\newcommand{\ketbra}[2]{\left|#1\right\rangle\!\!\left\langle #2\right|}

\def\eps{\varepsilon}

\def\id{\mathbb{I}}
\newcommand*{\Tr}{\mathrm{Tr}}

\def\D{\mathcal{D}}
\def\E{\mathcal{E}}
\def\A{\mathcal{A}}

\begin{document}

\title{Free entropy and quantum minimum description length}
%\author{ \\
%{\small \it Stanford Institute for Theoretical Physics, Stanford University, Stanford, CA 94305}}
\author{Patrick Hayden}
\affiliation{
 Leinweber Institute for Theoretical Physics, Stanford University, Stanford, CA 94305, USA}
 \author{Alexander Maloney}
\affiliation{Department of Physics, Syracuse University, Syracuse, NY 13244, USA}
\affiliation{Institute for Quantum and Information Sciences, Syracuse University, Syracuse, NY 13244, USA}
\author{Jinzhao Wang}
\email{wang.jinzhao226@gmail.com}
\affiliation{
 Leinweber Institute for Theoretical Physics, Stanford University, Stanford, CA 94305, USA}
\affiliation{Zyphra, San Francisco, CA 94105, USA}

\author{Yuxiang Yang}
\affiliation{
QICI Quantum Information and Computation Initiative, School of Computing and Data Science, The University of Hong Kong, Pokfulam Road, Hong Kong, China}
%date{}

\begin{abstract}
In free probability, Voiculescu's free entropy counts matrix microstates that approximate an operator. It is a non-commutative analog of the Shannon entropy, distinct from any quantum entropy of a density operator, but it has had no physical meaning in quantum theory. We define a finite-dimensional physical free entropy at finite resolution and show that it directly characterizes, up to an explicit constant, the minimal quantum memory needed to program many copies of a state with known eigenvalues and unknown eigenbasis. This task is a variant of Schumacher compression, and we call its optimal cost the quantum minimum description length. We briefly discuss how free entropy differs from quantum entropies and its potential applications in quantum physics.
\end{abstract}

\maketitle

\emph{Introduction.}---Quantum information theory is largely built on quantum entropies, such as the von Neumann entropy, much as classical information theory is built on classical entropies, such as the Shannon entropy. A familiar route from a classical entropy to a quantum one replaces a probability distribution by a density operator in an appropriate functional formula. For example, one goes from the Shannon entropy to the von Neumann entropy via $H( p)\equiv-\sum_i p_i\log p_i\rightarrow S(\rho) \equiv -\Tr \rho\log \rho.$ One can then look for operational meanings of these quantities in information-processing tasks.

Switching from a probability vector to a density operator is not the only way to a non-commutative counterpart of the Shannon entropy. A natural alternative starts from a different expression for the Shannon entropy,
\begin{equation}\label{eq:shannon}
    H({p})=\inf_{\eps>0}\lim_{N\to\infty}\frac1N\log \#\{ x^N\in \mathcal{X}^N \, \big\vert\, \| p_{x^N}-{p}\|_1\le\eps \}\ ,
\end{equation}
where $\mathcal{X}$ denotes a finite alphabet and $ p_{x^N}$ is the empirical type (frequency vector) that records the frequency of each symbol in the length-$N$ string $x^N$. What we count is the size of the \emph{typical set}. We use the binary logarithm throughout this Letter.

Equation~\eqref{eq:shannon} captures the essence of entropy: it is of the form $\log W$, with $W$ a number of ``microstates'' (typical strings), and it manifests its operational meaning as the optimal compression rate for data generated by the source $p$, \`a la Shannon's source coding theorem.

The von Neumann entropy $S(\rho) = -\Tr \rho\log \rho$ also admits a Boltzmann-type expression that underpins Schumacher's quantum data compression~\cite{schumacher1995quantum}:
\begin{equation}\label{eq:vN}
S(\rho)
=\inf_{0<\eps<1}\ \limsup_{N\to\infty} \frac1N
\min_{\substack{V\subset\mathcal H^{\otimes N}:\\
\Tr\rho^{\otimes N}\Pi_V\ge 1-\eps}}\log\dim V\ ,
\end{equation}
where $\rho^{\otimes N}$ is the $N$-copy state on $\mathcal H^{\otimes N}$, the minimum runs over subspaces $V\subset \mathcal H^{\otimes N}$, and $\Pi_V$ denotes the
orthogonal projector onto $V$.
The quantity inside the logarithm is the dimension of a subspace $V$ that captures
at least $1-\eps$ of the weight of $\rho^{\otimes N}$\footnote{Write $\rho=\sum_{i=1}^r p_i\ketbra{i}{i}$ on its support, with $p_i>0$. The smallest admissible $\dim V$ in~\eqref{eq:vN} is $\min_{A\subseteq\{1,\dots,r\}^N}\{\#A:\sum_{i^N\in A}\prod_{k=1}^N p_{i_k}\ge1-\eps\}$: one may choose the most likely product eigenvectors. Taking $A$ to be the strings whose empirical type is close to $p$ gives the direct count in~\eqref{eq:shannon} and the same asymptotic rate, though not necessarily the finite-$N$ minimum.}.

What if we quantize the Shannon entropy via~\eqref{eq:shannon} by promoting the microstates $x^N$ to be non-commutative, rather than resorting to density operators? \emph{Free entropy} was invented as such a non-commutative analogue of the Shannon entropy~\cite{voiculescu1993analogues,voiculescu1994analogues,voiculescu1996analogues,voiculescu1998analogues,voiculescu1999analogues,voiculescu2002free}. It generalizes~\eqref{eq:shannon} by promoting the strings $x^N$ to $N\times N$ matrices $X_N$. For a bounded self-adjoint operator $x$ in a tracial von Neumann algebra $(\A,\tau)$\footnote{The definition can also be adapted to certain non-self-adjoint operators, such as unitaries.}, fix an operator-norm cutoff $R>\|x\|$. Its free entropy $\chi(x)$ is defined as
\begin{multline}\label{eq:free_ent}
\chi(x)=\inf_m\inf_{\eps>0}\limsup_{N\to\infty}\bigg[\frac12\log N
+\frac1{N^2}\log\mathrm{Vol}\\
\left\{X_N\in M_N^{\mathrm{s.a.}}:
\begin{array}{l}
\|X_N\|\le R,\\
|N^{-1}\Tr X_N^k-\tau(x^k)|\le\eps\\
\text{for }1\le k\le m
\end{array}\right\}\bigg],
\end{multline}
where $M^{\mathrm{s.a.}}_N$ denotes the set of $N\times N$ self-adjoint matrices over complex numbers, and the microstates $X_N$ are required to reproduce \emph{all} moments up to order $m$ within tolerance $\eps$. The result is independent of the chosen $R>\|x\|$~\cite{voiculescu2002free}. We demand closeness in moments as the typicality condition\footnote{For a single variable one could instead compare spectral distributions in a weak metric. The $\ell_1$ distance used in~\eqref{eq:shannon} does not work here, because an atomic empirical measure is at maximal total-variation distance from any continuous one. Moments have the further advantage that they extend to several non-commuting variables, which have no joint spectral distribution.}, and we take a $\limsup$ because the limit need not exist. The number of microstates is measured by the Lebesgue volume (denoted Vol) on $M^{\mathrm{s.a.}}_N$, which the Hilbert--Schmidt inner product identifies with $\mathbb{R}^{N^2}$\footnote{A Gaussian reference measure instead produces a related free-energy functional containing an additional quadratic-moment term, so we keep the Lebesgue convention here.}. The normalization is adjusted to $1/N^2$ because a matrix carries $O(N^2)$ degrees of freedom. Hence, the definition is an analog of the Shannon entropy~\eqref{eq:shannon}, up to the term $\frac12\log N$. This term removes a universal divergence in the volume. As its name suggests, free entropy is the entropy adapted to \emph{free independence}: it is additive for freely independent families, just as the Shannon and von Neumann entropies are additive under tensor products~\cite{voiculescu1994analogues}, a point we return to later.

In fact, as we ``count'' using the Lebesgue measure, free entropy is more akin to Shannon's differential entropy,
\begin{equation}\label{eq:diff_ent_formula}
   h( p) \equiv - \int_K p( x)\log p( x)\,\dd x\ ,
\end{equation}
than the discrete Shannon entropy: for a distribution on a compact subset $K\subset\mathbb R^{k}$, $h(p)$ admits an entirely analogous microstate expression, in which one counts by Lebesgue volume the tuples $x^N\in K^N$ whose moments approximate those of $p$ (cf. Proposition 5.1.1 in~\cite{hiai2000semicircle}).

Free entropy~\eqref{eq:free_ent} generalizes this microstate expression by counting, instead of typical tuples, ``typical'' matrices whose empirical spectral distributions approximate that of $x$. For a single self-adjoint variable, $\chi(x)$ has the following closed-form expression in terms of its spectral measure $\mu_x$~\cite{voiculescu1994analogues,arous1997large,hiai2000semicircle}:
\begin{multline}\label{eq:free_ent_formula}
    \chi(x)=\iint_{\mathbb R^2}\log|s-t|\,\dd\mu_x(s)\,\dd\mu_x(t)\\
    +\frac34\log e+\frac12\log(2\pi).
\end{multline}
The additive constant depends on how the volume is normalized, so we focus on the logarithmic potential term. It plays the role that $-\int p\log p$ plays in~\eqref{eq:diff_ent_formula}.

The logarithmic potential has a simple origin. In eigenvalue and eigenvector coordinates, the Lebesgue measure is $\dd X\propto\Delta(\lambda)^2\,\dd\lambda\,\dd U$, where $\Delta(\lambda)=\prod_{i<j}(\lambda_i-\lambda_j)$ is the Vandermonde determinant. The microstates are constrained only through their spectra, so the eigenvectors contribute a normalization, and $N^{-2}\log\Delta(\lambda)^2=N^{-2}\sum_{i\ne j}\log|\lambda_i-\lambda_j|$ is a discretized logarithmic energy.

In mathematics, free entropy has been used as a tool to probe Type II von Neumann algebras~\cite{voiculescu1996analogues,voiculescu2002free,ge1997applications,ge1998applications}, which was the primary purpose of its invention. While free entropy also admits a large-deviation interpretation in random matrix theory~\cite{arous1997large}, to our knowledge no operational meaning in physics has been identified. It would be unsatisfactory if such a natural generalization of the Shannon entropy had no role in physics. The goal of this Letter is to find one. For a state with known, strictly positive, non-degenerate spectrum but unknown eigenbasis, the optimal asymptotic memory cost for programming $n$ copies is $|M_n|=\tfrac12\chi_{\mathrm{phy}}(\rho;n^{-1})+C_{d,d}+o(1)$, Eq.~\eqref{eq:result}, where $\chi_{\mathrm{phy}}$ is the finite-dimensional, finite-resolution free entropy introduced next. The End Matter extends this relation to rank-deficient states with distinct positive eigenvalues.

\emph{Physical free entropy.}---The theory of free entropy is usually formulated in infinite dimensions for random variables in a tracial von Neumann algebra, such as a Type~II$_1$ algebra. To find its role in quantum theory, it is convenient to put it in a more standard finite-dimensional setting. In finite dimensions, we have $\rho\in (M_d(\mathbb{C}),d^{-1}\Tr)$, and $\mu_\rho = \frac1d\sum_{i=1}^d\delta_{p_i}$ is an atomic spectral measure at the eigenvalues $p_i$ of $\rho$. In the operational task below, we focus on a density operator $\rho$. The definition also applies to any finite-dimensional self-adjoint operator.

However, if we substitute the atomic measure $\mu_\rho$ into the free entropy formula~\eqref{eq:free_ent_formula}, we obtain $\chi(\rho)=-\infty$: every atom has a divergent self-interaction. The microstate definition~\eqref{eq:free_ent} is applicable, but it records these atomic self-collisions rather than a finite-dimensional orbit volume. When $N>d$, a larger matrix $X_N$ also possesses more degrees of freedom than are needed to describe the smaller $d$-dimensional $\rho$. This is a problem of large $N$.

Another issue is that free entropy can be negative. This is not a problem mathematically, but a negative entropy would be difficult to interpret physically. It comes from measuring microstates with a continuous volume.

We resolve both issues by counting at a finite resolution $\eps$ at a fixed $N=d$. Order the eigenvalues as $p_1\ge\cdots\ge p_d$ and $\lambda_1(X)\ge\cdots\ge\lambda_d(X)$, and set
\[
\Omega_\eps=\{X\in M_d^{\mathrm{s.a.}}:\|\lambda(X)-p\|_2\le\eps\}.
\]
We define the \emph{physical free entropy} by
\begin{equation}\label{eq:free_phys_def}
    \chi_\mathrm{phy}(\rho;\eps)=\log\frac{\mathrm{Vol}(\Omega_\eps)}{\mathrm{Vol}(B_\eps^{d^2})},
\end{equation}
where $B_\eps^{d^2}$ is a Hilbert--Schmidt ball of radius~$\eps$. The Hoffman--Wielandt inequality bounds the distance below by the distance between the ordered eigenvalue lists~\cite{hoffman1953variation}. Aligning their eigenbases attains the bound, so
\[
\min_{U\in\mathrm U(d)}\|X-U\rho U^\dagger\|_2=\|\lambda(X)-p\|_2,
\]
so $\Omega_\eps$ is precisely the $\eps$-neighborhood of the unitary orbit in the full Hermitian matrix space. It can include matrices that are not density operators. In particular, $\Omega_\eps$ contains the ball of radius $\eps$ around $\rho$, so $\chi_\mathrm{phy}\ge0$, with equality only when $\rho$ is proportional to the identity. The volume ratio gives an effective count of resolution cells, with this ambient-space convention fixing its additive constant\footnote{At fixed $d$, its logarithm agrees with the logarithm of the covering number up to $O(1)$ as $\eps\to0$~\cite{Kolmogorov1963,KolmogorovTikhomirov1961}. We use the volume ratio to fix that ambiguity.}.

Let $g$ be the smallest gap between distinct eigenvalues, with $g=\infty$ for a scalar operator, and take $\eps<g/2$. The End Matter shows that
\begin{multline}\label{eq:free_phys}
    \chi_\mathrm{phy}(\rho;\eps)=N_\rho\log\eps^{-1}+\chi_\mathrm{reg}(\rho)
    +\mathrm{const}+O_d(\eps^2/g^2),
\end{multline}
where $N_\rho=d^2-\sum_a g_a^2$ is the real orbit dimension, $g_a$ are the spectral multiplicities, the constant depends only on these multiplicities and $d$, and
\begin{equation}\label{eq:reg_free_ent}
    \chi_\mathrm{reg}(\rho)=2\sum_{i<j,\,p_i\ne p_j}\log|p_i-p_j|.
\end{equation}
Thus $d^{-2}\chi_\mathrm{reg}$ is the logarithmic energy in~\eqref{eq:free_ent_formula} with coincident-eigenvalue terms omitted. The leading coefficient also has a free-probability interpretation: $N_\rho/d^2$ is Voiculescu's free entropy dimension for the atomic measure $\mu_\rho$~\cite{voiculescu2002free}.

The main text treats full-rank states with distinct eigenvalues, for which $N_\rho=d^2-d$ and the expansion becomes
\begin{multline}\label{eq:free_phys1}
    \chi_\mathrm{phy}(\rho;\eps)=(d^2-d)\log\eps^{-1}
    +2\sum_{i<j}\log|p_i-p_j|\\
    +\mathrm{const}+O_d(\eps^2/g^2).
\end{multline}
The first term is determined by the real dimension of the fixed-spectrum unitary orbit, and the second term refines its volume. This is the free entropy quantity we shall find an operational meaning for. It also recovers Voiculescu's free entropy as a double scaling limit in the dimension and the resolution (Remark~\ref{rem:bridge} in the End Matter).

We should think of the physical free entropy as a direct analog of the Shannon entropy itself. Discretizing the Lebesgue measure used in the free entropy at resolution $\eps$ turns the differential entropy, which can be negative, into a Shannon entropy, $H(p_\eps)\simeq k\log\eps^{-1}+h(p)$ (cf.\ Chapter 8 of~\cite{cover1999elements}). Likewise, the volume ratio in~\eqref{eq:free_phys_def} gives an effective count of matrix microstate cells at resolution~$\eps$~\cite{kolmogorov1956shannon,renyi1959dimension,jaynes1957information}.

\begin{figure}[t]
    \includegraphics[width=\linewidth]{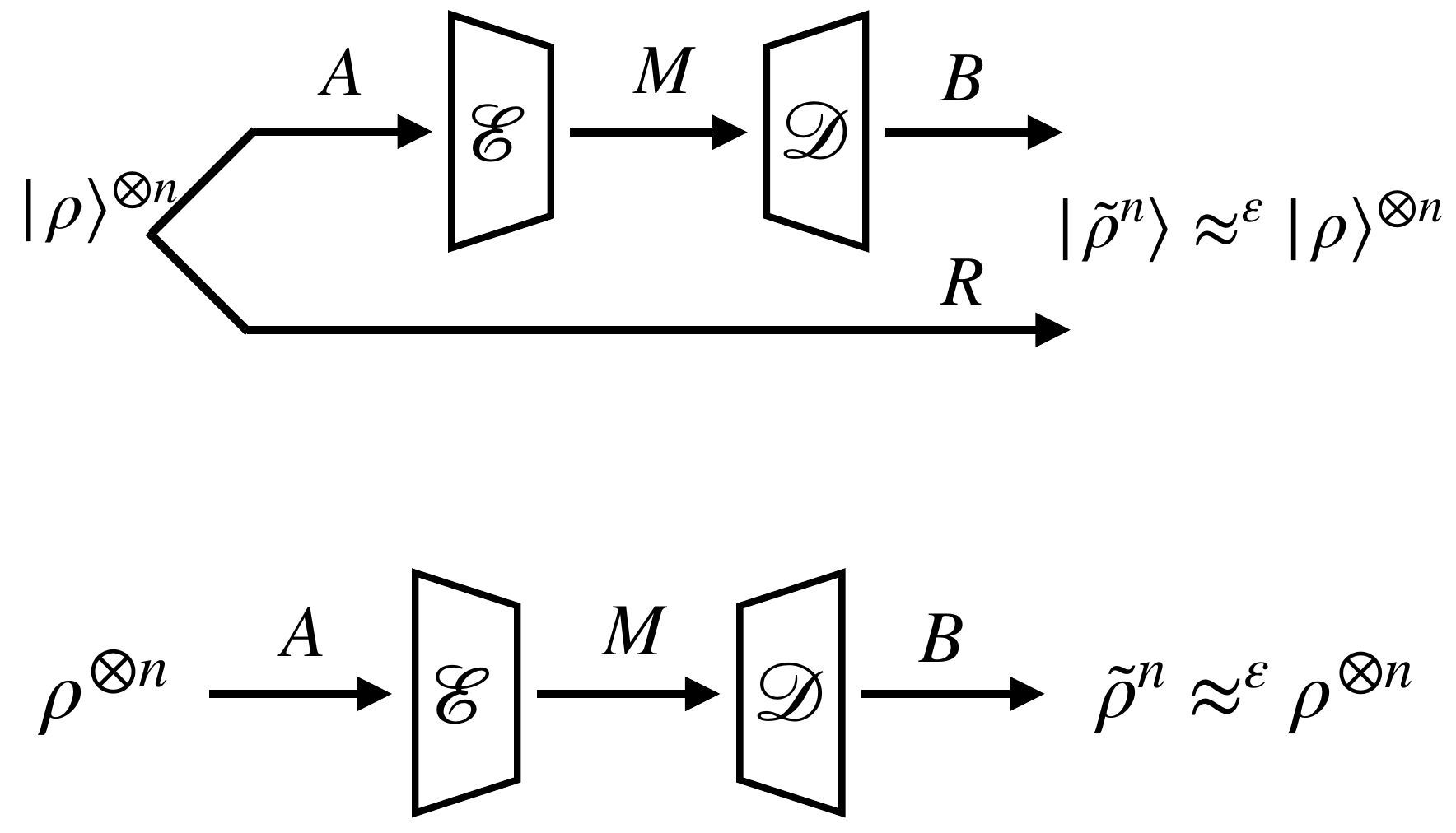}
    \caption{\label{fig:compression}\textbf{Schumacher compression (top) and our compression task (bottom).} The key distinction is that Schumacher compression must preserve a purification of $\rho^{\otimes n}$ with the reference system $R$. In our task the spectrum of the source is known but its eigenbasis is not, and only the $n$-copy state itself has to be recovered.}
\end{figure}

\emph{Compression.}---Entropy in information theory is always tied to compression. This is the case for Shannon entropy and von Neumann entropy. We therefore seek a quantum compression task whose minimal memory size is characterized by free entropy.

Fix $p_1>\cdots>p_d>0$ and write $\rho_U=U\rho_0U^\dagger$, where $\rho_0=\mathrm{diag}(p_1,\ldots,p_d)$.
An encoder $\E_n$ maps $\rho_U^{\otimes n}$ into a memory $M_n$, and a decoder $\D_n$ reconstructs the state. Both maps are quantum channels that may depend on $n$ and the spectrum, but not on $U$. The memory includes every retained quantum or classical register, with cost $|M_n|=\log\dim M_n$\footnote{We use log dimension. Rounding up to a whole number of qubits can change the additive constant.}. We require vanishing \emph{global} error,
\begin{equation}\label{eq:error}
    \delta_n=\sup_{U\in\mathrm U(d)}\frac12
    \left\|\rho_U^{\otimes n}-\D_n\circ\E_n(\rho_U^{\otimes n})\right\|_1
    \longrightarrow0.
\end{equation}
The optimal asymptotic memory cost over code sequences satisfying this condition is the quantity we compare with free entropy.

Tomography would give a classical description of $\rho_U$. Here we want instead a minimal \emph{quantum} description $\E_n(\rho_U^{\otimes n})$ from which $\rho_U^{\otimes n}$ can be programmed. We therefore call the optimal memory cost the \emph{quantum minimum description length} (QMDL) of the known-spectrum orbit family, by analogy with the classical minimum description length~\cite{rissanen1978modeling}.

Note that the error~\eqref{eq:error} is evaluated on the whole $n$-copy state. For this reason one cannot simply estimate $\rho_U$ and prepare $n$ copies of the estimate: measuring and re-preparing always leaves a global error that does not vanish~\cite{yang2018compression}.

For fixed dimension and spectrum, a vanishing-error code sequence attains
\begin{multline}\label{eq:result_qmdl}
    |M_n|=\frac{d^2-d}{2}\log n+\sum_{i<j}\log|p_i-p_j|\\
    -\sum_{k=1}^{d-1}\log k!+o(1).
\end{multline}
Let $L_n(p)$ denote the right-hand side of~\eqref{eq:result_qmdl} without $o(1)$. Every vanishing-error code sequence obeys $\liminf_{n\to\infty}(|M_n|-L_n(p))\ge0$. The companion paper~\cite{paper2} gives the construction and converse. Here we explain their connection to free entropy.

For the attaining sequence, comparison with~\eqref{eq:free_phys1} and its constant in~\eqref{eq:free_phys1_app} gives our central relation,
\begin{equation}\label{eq:result}
    |M_n|=\frac12\chi_\mathrm{phy}(\rho;n^{-1})+C_{d,d}+o(1),
\end{equation}
where the spectrum-independent offset is
\begin{multline}\label{eq:result_const}
    C_{d,d}=-\frac12\sum_{k=1}^{d-1}\log k!
    -\frac12\log\frac{\Gamma(d^2/2+1)}{\Gamma(d/2+1)}\\
    -\frac{d(d-1)}4\log2.
\end{multline}
That the equality holds only up to a $d$-dependent, state-independent constant is expected. Free entropy measures a continuous volume, so its additive offset reflects the chosen metric and reference-cell convention, just as for differential entropy. The state-dependent part, driven by eigenvalue repulsion, is captured exactly. Physical free entropy is unnormalized in~\eqref{eq:free_phys_def}. The factor $d^{-2}$ enters only when taking the large-dimension limit.

The form of~\eqref{eq:result_qmdl} comes from the representation retained by the attaining code. Schur--Weyl duality gives $\rho_U^{\otimes n}=\bigoplus_\lambda\pi_\lambda(\rho_U)\otimes\id_{\mathcal K_\lambda}$, where $\pi_\lambda$ acts on the irrep $\mathcal H_\lambda$ and $\mathcal K_\lambda$ is the permutation-group multiplicity space. The factor on $\mathcal K_\lambda$ does not depend on $U$ and can be regenerated, while the Young diagram concentrates around $\lambda/n\approx p$~\cite{keyl2001estimating}. The protocol of~\cite{paper2} maps typical blocks into one target irrep $\mathcal H_\Lambda$ with $\Lambda/n\to p$, so its memory has dimension $\dim\mathcal H_\Lambda$. Weyl's dimension formula gives~\cite{fulton1991representation}
\begin{equation}\label{eq:weyl_geometry}
\dim\mathcal H_\Lambda=[1+o(1)]\,
\frac{n^{d(d-1)/2}\Delta(p)}{\prod_{k=1}^{d-1}k!},
\end{equation}
where $\Delta(p):=\prod_{i<j}(p_i-p_j)$. Taking its logarithm gives the $\log n$, eigenvalue-gap, and factorial terms in~\eqref{eq:result_qmdl}. 

The factor of $1/2$ in~\eqref{eq:result} reflects two volume forms on the unitary orbit. Tangent vectors at $\rho$ are commutators $[K,\rho]$, with $K$ anti-Hermitian, and in the eigenbasis of $\rho$,
\[
\bra{i}[K,\rho]\ket{j}=(p_j-p_i)\,K_{ij}\ ,\qquad K_{ij}=\bra{i}K\ket{j}\ .
\]
The induced Hilbert--Schmidt volume weights both real directions of $K_{ij}$ by the eigenvalue gap and therefore carries $\Delta(p)^2$.  A representation dimension is instead a symplectic volume~\cite{kirillov2004orbit}. In the Kirillov--Kostant--Souriau (KKS) symplectic form, these directions form one canonical pair and contribute one gap to the symplectic volume, giving $\Delta(p)$.

The choice $\eps=n^{-1}$ is a resolution in the volume comparison, not an estimation precision for each basis parameter. With $n$ copies, neighboring states on the orbit become distinguishable at angular scale $n^{-1/2}$, consistent with the $\frac12\log n$ cost per real orbit direction.

For the spectra treated here, this result provides an operational meaning for free entropy, in parallel with Schumacher compression giving an operational meaning for von Neumann entropy~\cite{schumacher1995quantum} (cf. Fig.~\ref{fig:compression}). The key distinction is that the error~\eqref{eq:error} is judged without including the purification, whereas Schumacher compression preserves it. Moreover, in our task, the state is unknown to the encoder and decoder except for its known spectrum, whereas the source state is usually assumed known in Schumacher compression\footnote{The version where the state is unknown is called ``universal quantum data compression''~\cite{jozsa1998universal,hayashi2002quantum,hayashi2002simple,hayashi2010universal}.}. 

Pure states make the contrast sharp. For a pure state $\psi$ there is no purification to preserve, so the two tasks differ only in what the encoder knows. The von Neumann entropy of $\psi$ is zero, and Schumacher compression is trivial when $\psi$ is known. The physical free entropy of $\psi$ is not zero (see the End Matter), and an observer who does not know $\psi$ needs a memory of $(d-1)\log n-\log(d-1)!+o(1)$ qubits to store $\psi^{\otimes n}$ for all pure states at once.

Earlier work established the leading memory cost~\cite{yang2016optimal,yang2018compression}, and the qubit protocol of~\cite{yang2016optimal} already attains a spectrum-dependent constant. The relation~\eqref{eq:result} identifies the free-entropy content of the optimal cost through order one.

One application of programming states is the storage of finite quantum reference frames used in covariant estimation and programming (cf.\ the quantum stopwatch~\cite{yang2018quantum} and optimal programming~\cite{yang2020optimal,Kubicki_2019}). 
 
What about compressing states with an unknown spectrum? To leading order the answer is $|M_n| = \frac{d^2-1}{2}\log n+o(\log n)$~\cite{yang2016optimal,yang2018compression}. We expect the order-one term to depend only on $d$, which makes it less interesting here. The leading coefficient again has a counting interpretation, but now at the level of the \emph{family} of states rather than of any single operator: $N_\rho$ is a property of the operator, not of the observer's knowledge ($N_\rho=d^2-d$ for generic $\rho$), whereas the family of all states carries $d^2-1$ real parameters. Its metric entropy at the local distinguishability scale $n^{-1/2}$ is therefore $\frac{d^2-1}{2}\log n$ at leading order. Equivalently, this is half the formal metric-entropy expression evaluated at scale $n^{-1}$, in parallel with~\eqref{eq:result}. Connecting it to free entropy would require a Bayesian framework to model the empirical spectral distribution for the whole family, which we do not attempt to spell out here.

\emph{Discussion.}---The QMDL suggests an analogy with the quantum Kolmogorov complexity of the known-spectrum orbit family of $n$-copy states~\cite{mueller2007,Berthiaume_2001,G_cs_2001}, although we do not claim an equality with any machine-dependent notion of individual description length. Classically, a string consisting of a fixed repeated pattern has description length $K(n)+O(1)=O(\log n)$: one records its length and the pattern. Our result has a similar structure, with universal logarithmic scaling and the state dependence in the order-one term. Unlike a classical pattern, however, the eigenbasis cannot be recorded exactly: $n$ copies resolve it only to the statistical scale $n^{-1/2}$, which costs $\frac12\log n$ per orbit degree of freedom.

QMDL also appears in \emph{universal lossless quantum compression}~\cite{hayashi2010universal,clarke1990information}. A source coding scheme that must work for every eigenbasis pays an extra average length above the familiar $nS(\rho)$. For the known-spectrum family, the smallest worst-case excess in ideal average length equals the QMDL up to a finite, spectrum-dependent entropy correction~\cite{paper2}. Physical free entropy therefore also describes the coding cost of an unknown eigenbasis. 

% Free entropy vs vN entropy
We would like to close by highlighting two important distinctions of free entropy from quantum entropy. 

Firstly, the von Neumann entropy is a functional of a state, conventionally represented by a density operator when an ambient trace is available. Voiculescu's microstate free entropy instead takes a tracial non-commutative distribution (a collection of moments) as its input. It therefore applies not only to self-adjoint variables but also to unitaries and families of observables\footnote{Standard microstate free entropy does not directly apply to type~III algebras, where no trace exists. A non-tracial extension requires separate treatment.}.

Within settings where free entropy is defined, we conjecture that it governs the QMDL of more general operators, through a relation of the same form as~\eqref{eq:result}. The relevant task is \emph{quantum programming}: how much quantum memory is needed to store a program for a family of operators, with error vanishing as the memory grows. We establish this connection here for states whose nonzero eigenvalues are distinct. Programming has also been studied for unitaries and measurements at leading order~\cite{Kubicki_2019,yang2020optimal,DAriano_2005,PrezGarca_2006}. We expect free entropy to appear in the order-one term of the program size more generally.

Secondly, ``free'' refers to \emph{freeness}, or free independence, a notion of independence with no classical analog~\cite{speicher1996universal}: alternating products of centered operators from free algebras have vanishing trace, so their mixed moments are fixed by their separate distributions. Free entropy is additive for freely independent families, as the von Neumann entropy is additive under tensor products~\cite{voiculescu1994analogues}. Physically, in suitable large-system or large-$N$ chaotic limits, observables separated in time can become asymptotically free~\cite{vardhan2025}, whereas tensor independence usually arises from spatial locality.

We have only treated the single-variable free entropy here. The multivariate theory is much richer, and we expect multivariate free entropy to govern the QMDL of several operators given all of their relations. This fits with Voiculescu's subadditivity, $\chi(x_1,\dots,x_k)\le\sum_i\chi(x_i)$, with equality for freely independent families~\cite{voiculescu1994analogues}. Describing the operators one at a time should never be cheaper.

Moreover, one can then naturally define the \emph{free mutual information} (FMI) and ask if it has any operational relevance in a communication task. A recent work by one of us uses FMI as a mutual information across time to probe quantum chaos and quantify how scrambling a unitary channel is, even though all unitary channels preserve information and preserve quantum entropies~\cite{vardhan2025}. We therefore expect free entropic quantities, like FMI, to be useful in capturing quantum information in quantum dynamics in a way that quantum entropies could not.

Overall, free entropy offers a perspective on quantum information complementary to that of the quantum entropies. Given their strong parallels, we envision that a new kind of information theory remains to be developed.

\par\noindent
\emph{Acknowledgments.}---YY acknowledges the support of the National Natural Science Foundation of China via the Excellent Young Scientists Fund (Hong Kong and Macau) Project 12322516 and the support of the Research Grants Council (RGC) of Hong Kong SAR via the NSFC/RGC Joint Research Scheme N\_HKU7107/24.
This material is based upon work supported by the U.S. Department of Energy, Office of Science, Office of High Energy and Nuclear Physics, under award number DE-SC0026551. The work of AM is supported by the Simons Foundation Grant No. 12574.
\par\noindent
\emph{Data availability.}---No data were generated or analyzed.

\par\noindent
\emph{AI Usage.}---All the ideas and results in this Letter were due to the authors. AI tools were used to assist with the text revision and editing process.
\bibliography{free}

\section*{End Matter}
\appendix
\setcounter{secnumdepth}{1}

\section{Derivation of the physical free entropy formula}

First suppose that $p_1>\cdots>p_d$. In eigenvalue--eigenvector coordinates, ordering the eigenvalues removes the permutation ambiguity, and normalized invariant measure $\dd\nu(U)$ on $\mathrm U(d)/\mathrm U(1)^d$ accounts for the diagonal-phase stabilizer. The Hilbert--Schmidt volume element is~\cite{zyczkowski2003hilbert}
\begin{equation}\label{eq:hs_volume_element}
    \dd X=\frac{(2\pi)^{d(d-1)/2}}{\prod_{k=1}^{d-1}k!}
    \Delta(\lambda)^2\,\dd\lambda_1\cdots\dd\lambda_d\,\dd\nu(U).
\end{equation}
For $\eps<g/2$, the eigenvalue ball $B_\eps^d(p)$ stays inside the ordered chamber. Taylor expansion of $\Delta(\lambda)^2$ about $p$ has no linear contribution after integration over this ball, so
\begin{multline}
    \mathrm{Vol}(\Omega_\eps)
    =\frac{(2\pi)^{d(d-1)/2}}{\prod_{k=1}^{d-1}k!}
    \frac{\pi^{d/2}\eps^d}{\Gamma(d/2+1)}
    \prod_{i<j}(p_i-p_j)^2\\
    {}\times\left[1+O_d\!\left(\frac{\eps^2}{g^2}\right)\right].
\end{multline}
Dividing by $\mathrm{Vol}(B_\eps^{d^2})=\pi^{d^2/2}\eps^{d^2}/\Gamma(d^2/2+1)$ and taking the logarithm gives
\begin{multline}\label{eq:free_phys1_app}
    \chi_{\mathrm{phy}}(\rho;\eps)
    =(d^2-d)\log\eps^{-1}
    +2\sum_{i<j}\log|p_i-p_j|\\
    -\sum_{k=1}^{d-1}\log k!
    +\log\frac{\Gamma(d^2/2+1)}{\Gamma(d/2+1)}
    +\frac{d(d-1)}2\log2\\
    +O_d\!\left(\frac{\eps^2}{g^2}\right).
\end{multline}
For a qubit with fixed spectrum $(p,1-p)$, $p>1/2$, this reads
\[
\frac12\chi_\mathrm{phy}(\rho;n^{-1})
=\log n+\log(2p-1)+1+O(n^{-2}).
\]
The QMDL is $\log n+\log(2p-1)+o(1)$~\cite{paper2}, consistent with $C_{2,2}=-1$ in~\eqref{eq:result_const}.

\begin{rem}\label{rem:bridge}
Physical free entropy recovers Voiculescu's $\chi$ as a double scaling limit in the dimension and the resolution. Let $\mu_x$ be compactly supported with finite logarithmic energy, and let $X_d$ be a self-adjoint matrix with decreasingly ordered eigenvalues $a_{1,d}>\cdots>a_{d,d}$ given by the quantiles of $\mu_x$. Assume that these quantile discretizations have nonzero gaps and convergent discrete logarithmic energies, as holds under standard regularity conditions (for example, for a continuous density bounded above and below on a compact interval). Then for any $\eps_d\to0$ with $\eps_d=o(g_d)$ and $\log\eps_d^{-1}=o(d)$, where $g_d$ is the minimal spectral gap of $X_d$,
\begin{multline}\label{eq:bridge}
\chi(x)=\lim_{d\to\infty}\Big[
\tfrac{1}{d^{2}}\,\chi_{\mathrm{phy}}(X_d;\eps_{d})
-\log\eps_{d}^{-1}-\tfrac12\log d\Big]\\
+\tfrac12\log e+\tfrac12\log(2\pi)\ .
\end{multline}
\end{rem}

\emph{Derivation of Remark~\ref{rem:bridge}.} Applying Stirling's formula to the factorial and gamma terms in~\eqref{eq:free_phys1_app} gives
\begin{multline}
\frac1{d^2}\left[-\sum_{k=1}^{d-1}\log k!
+\log\frac{\Gamma(d^2/2+1)}{\Gamma(d/2+1)}
+\frac{d(d-1)}2\log2\right]\\
=\frac12\log d+\frac14\log e+o(1).
\end{multline}
Consequently,
\begin{multline}
\frac1{d^2}\chi_\mathrm{phy}(X_d;\eps_d)
-\log\eps_d^{-1}-\frac12\log d
=-\frac1d\log\eps_d^{-1}\\
+\frac2{d^2}\sum_{i<j}\log|a_{i,d}-a_{j,d}|
+\frac14\log e+o(1).
\end{multline}
To justify the remainder as $d$ varies, use $\eps_d/g_d\to0$: throughout the eigenvalue ball, the logarithm of each Vandermonde factor changes by $o(1)$ uniformly. Summing over $O(d^2)$ factors and dividing by $d^2$ leaves $o(1)$, without needing a uniform Taylor constant. The resolution term vanishes by $\log\eps_d^{-1}=o(d)$, and the double sum converges to the logarithmic energy of $\mu_x$ by assumption. Comparison with~\eqref{eq:free_ent_formula} proves~\eqref{eq:bridge}.

\section{Free entropy of density operators with a degenerate spectrum}
If the spectrum is degenerate, let its distinct eigenvalues be $p_1>\cdots>p_m$ with multiplicities $g_1,\ldots,g_m$ and $\sum_a g_a=d$. Define
\[
    \kappa:=\sum_a g_a^2,\qquad
    N_\rho:=d^2-\kappa,\qquad
    g:=\min_{a<b}(p_a-p_b),
\]
with $g=\infty$ when $m=1$. For $\eps<g/2$, different eigenvalue blocks cannot cross. Writing $\lambda_{a,r}=p_a+x_{a,r}$, the integration domain is therefore
\[
\sum_{a,r}x_{a,r}^2\le\eps^2,\qquad
x_{a,1}\ge\cdots\ge x_{a,g_a}\quad\text{for each }a.
\]
The squared Vandermonde factorizes into inter-block and intra-block parts,
\begin{multline*}
\Delta(\lambda)^2=\prod_{a<b}\prod_{r=1}^{g_a}\prod_{s=1}^{g_b}\big(p_a-p_b+x_{a,r}-x_{b,s}\big)^2\\
\times\prod_{a=1}^m\prod_{1\le r<s\le g_a}(x_{a,r}-x_{a,s})^2\ ,
\end{multline*}
and the inter-block part equals $\prod_{a<b}(p_a-p_b)^{2g_ag_b}\,[1+L(x)+O_d(\eps^2/g^2)]$ with $L$ linear in $x$. The map $x_{a,r}\mapsto-x_{a,g_a+1-r}$ preserves the domain and the intra-block factors and flips the sign of $L$, so $L$ integrates to zero. Rescaling $x=\eps y$ extracts $\eps^{\,\sum_a g_a(g_a-1)}$ from the intra-block Vandermondes and $\eps^{d}$ from the eigenvalue measure, $\eps^{\kappa}$ in total. The remaining integral is
\begin{multline*}
\int_{\substack{\sum_{a,r}y_{a,r}^2\le1\\
 y_{a,1}\ge\cdots\ge y_{a,g_a}\ \forall a}}
\prod_a\prod_{r<s}(y_{a,r}-y_{a,s})^2\,\dd y\\
=\frac{(2\pi)^{d/2}\prod_a\prod_{k=1}^{g_a-1}k!}
{2^{\kappa/2}\Gamma(\kappa/2+1)}.
\end{multline*}
Indeed, integrating against $e^{-\|y\|_2^2/2}$ factorizes over the blocks. Applying~\eqref{eq:hs_volume_element} in each block evaluates the Gaussian integrals, and radial integration gives the displayed ball integral. Combining it with~\eqref{eq:hs_volume_element} yields
\begin{multline}\label{eq:degenerate_volume}
    \frac{\mathrm{Vol}(\Omega_\eps)}
    {\mathrm{Vol}(B_\eps^{d^2})}
    =
    2^{N_\rho/2}
    \frac{\Gamma(d^2/2+1)}{\Gamma(\kappa/2+1)}
    \frac{\prod_a\prod_{k=1}^{g_a-1}k!}
    {\prod_{k=1}^{d-1}k!}\\
    {}\times\eps^{-N_\rho}
    \prod_{a<b}(p_a-p_b)^{2g_ag_b}
    \left[1+O_d\!\left(\frac{\eps^2}{g^2}\right)\right].
\end{multline}
Consequently, the physical free entropy is
\begin{multline}\label{eq:degenerate_free_ent}
    \chi_\mathrm{phy}(\rho;\eps)
    =N_\rho\log \eps^{-1}
    +2\sum_{a<b}g_a g_b\log|p_a-p_b|\\
    +\frac{N_\rho}{2}\log2
    +\log\frac{\Gamma(d^2/2+1)}{\Gamma(\kappa/2+1)}\\
    +\sum_a\sum_{k=1}^{g_a-1}\log k!
    -\sum_{k=1}^{d-1}\log k!
    +O_d(\eps^2/g^2).
\end{multline}
The second term is $\chi_\mathrm{reg}(\rho)$ from~\eqref{eq:reg_free_ent}, while the remaining order-one terms depend only on $d$ and the multiplicities. When all $g_a=1$, this reduces to~\eqref{eq:free_phys1_app}.

Now let $\rho$ have rank $r<d$ and distinct positive eigenvalues $p_1>\cdots>p_r>0$. The zero eigenvalue has multiplicity $d-r$, while all positive eigenvalues are simple, so $N_\rho=r(2d-r-1)$ and $\kappa=(d-r)^2+r$. Equation~\eqref{eq:degenerate_free_ent} becomes
\begin{multline}
    \chi_\mathrm{phy}(\rho;\eps)
    = N_\rho\log \eps^{-1} +2\sum_{1\le i<j\le r}\log |p_i-p_j|
    \\+2(d-r)\sum_{i=1}^{r}\log p_i
    +\frac{N_\rho}2\log2
    +\log\frac{\Gamma(d^2/2+1)}
    {\Gamma(\kappa/2+1)}\\
    -\sum_{k=d-r}^{d-1}\log k!
    +O_d(\eps^2/g^2).
\end{multline}
For these spectra, an asymptotically optimal code sequence attains~\cite{paper2}
\begin{multline}\label{eq:rank_def_qmdl}
    |M_n|
    =\frac{N_\rho}2\log n
    +\sum_{1\le i<j\le r}\log|p_i-p_j|\\
    +(d-r)\sum_{i=1}^r\log p_i
    -\sum_{k=d-r}^{d-1}\log k!+o(1)\ .
\end{multline}
Thus, for any such rank-deficient $\rho$, the attaining sequence satisfies
\begin{equation}
    |M_n| = \frac{1}{2}\chi_\mathrm{phy}(\rho;n^{-1})
    + C_{d,r} + o(1)\ ,
\end{equation}
where the order-one offset is explicit:
\begin{multline}
    C_{d,r}
    =-\frac12\sum_{k=d-r}^{d-1}\log k!
    -\frac{N_\rho}4\log2\\
    -\frac12\log
    \frac{\Gamma(d^2/2+1)}{\Gamma(\kappa/2+1)}.
\end{multline}
For $r=d$, this reduces to~\eqref{eq:result_const}.

For a pure state, $r=1$ and $p_1=1$, so
\begin{multline}
    \chi_\mathrm{phy}(\ketbra{\psi}{\psi};\eps)
    =(2d-2)\log \eps^{-1}
    +(d-1)\log2\\
    {}-\log(d-1)!+\log\Gamma(d^2/2+1)\\
    -\log\Gamma\bigl(((d-1)^2+3)/2\bigr)+O_d(\eps^2).
\end{multline}
Here the orbit has dimension $2d-2$, and every nonzero eigenvalue gap equals one.

For the pure-state orbit family generated by $\ketbra{\psi}{\psi}$, the attaining memory cost is
\begin{equation}
    |M_n|=(d-1)\log n-\log(d-1)!+o(1),
\end{equation}
in agreement with the logarithm of the symmetric-subspace dimension $\binom{n+d-1}{d-1}$.

We expect that, with a more careful analysis for arbitrary spectral multiplicities, the QMDL of the known-spectrum orbit family generated by any density operator $\rho$ can be shown to equal $\frac12\chi_\mathrm{phy}(\rho;n^{-1})$ plus a multiplicity-dependent order-one offset and $o(1)$.
\end{document}